# A CURRENT-VOLTAGE CHARACTERISTIC OF PHOTORESISTANCE. A PLANE CASE

**Dimitar G. STOYANOV**

Faculty of Engineering and Pedagogy in Sliven, Technical University of Sofia
59, Bourgasko Shaussee Blvd, 8800 Sliven, BULGARIA

e-mail: dgstoyanov@abv.bg

**Abstract:** *The formation of photo-electron current in the volume of semi-conductor material is investigated in this article. A plane case when a material is uniformly illuminated by light is considered. The current-voltage characteristic of a photo resistance is obtained in analytical form.*

### 1. Introduction

The photoconductivity is an important characteristic of semi-conducting materials. At this moment of electronics development the photo processes in semi-conductors and semi-conducting details find more and more great practical application. Because of that the necessity of increasing the volumes of studying of these processes in books [1, 3] gets pressing.

The objective of this article is the analysis of the processes taking place in the formation of photoelectrons flow and its running. A plane case when a semi-conducting material is uniformly illuminated by light is considered.

### 2. Balance of Particles
### 2.1 Basic Processes

When the surface of semi-conducting material is illuminated by light the last penetrates in material volume, and as a result of an internal photo effect causes the appearance of electrons in the conduction band of semi-conductor. These electrons we will call photo electrons in order to distinguish them from the electrons in the conduction band which could be from the main current carriers or from material doping.

The generation of photo electrons in a unit of volume $N$ per a unit of time is proportional to the intensity of monochromatic light $I_\lambda$ [1, 2]. The proportionality coefficient $k_\lambda$ gives an account of the spectral sensitivity of the material along the wave's lengths $\lambda$ of the falling down light.

$$\left(\frac{\delta N}{\delta t}\right)_{ph} = + k_\lambda . I_\lambda \qquad (1)$$

The basic mechanism of photo electrons loss is their recombination with vacancies $N_p$ from the valence band. At small light intensity the number of vacancies generated from the photo effect is significantly smaller than the number of vacancies of the basic and non-basic current carriers. Because of that a great part of the process of recombination will set in with vacancies which concentration will not depend on light. Therefore, we can introduce a peculiar time of life to photo electrons $\tau$, which is a constant independent on light intensity magnitude [1, 2].

$$\left(\frac{\delta N}{\delta t}\right)_{rec} = -\gamma . N . N_p = -\frac{N}{\tau} \qquad (2)$$

On account of the small light intensity the photo electrons will not influence the intensity of the electrical field, and hence the Gauss' law will be executed

$$div\ \vec{E} = 0 \tag{3}$$

**2.2 Balance of Particles**

According to our opinion the balance of photo electrons should be made using the equation of continuity in which right side we put the sum of velocities from (1) and (2)

$$\frac{\partial N}{\partial t} + div\vec{\Gamma} = +k_\lambda . I_\lambda - \frac{N}{\tau}. \tag{4}$$

Here in equation (4) the quantity $\vec{\Gamma}$ figures, which the flow of photo electrons, generated by the presence of electric field with intensity $\vec{E}$. We will suppose that the flow velocity coincides with the drifted velocity of photo electrons $\vec{u}_d$

$$\vec{u}_d = -\mu.\vec{E}. \tag{5}$$

Here in equation (5) with $\mu$ is denoted the photo electrons mobility in the material ($\mu > 0$). The very flow of photo electrons will be of the form

$$\vec{\Gamma} = N.\vec{u}_d = -N.\mu.\vec{E}. \tag{6}$$

The flow of photo electrons will form a current of photo electrons, which we will call photo-current, with density $\vec{j}_f$

$$\vec{j}_f = -q.\vec{\Gamma} = q.N.\mu.\vec{E}. \tag{7}$$

Here $q$ is the magnitude of electron charge ($q > 0$). Replacing (6) in (4) we get

$$\frac{\partial N}{\partial t} + div(-N.\mu.\vec{E}) = +k_\lambda . I_\lambda - \frac{N}{\tau}. \tag{8}$$

This is the equation of the balance of photo electrons at the most general case. This can be solved together with equation (3) at the following limiting conditions:
a) The concentration and the flow of photo electrons is zero upon the negative end of the material;
b) The concentration and the flow of photo electrons are maximal upon the positive end of the material. The flow of photo electrons causes the photo-current $i_f$ read through the material.

**3. Plane Sample**
**3.1 Geometry of Problem**

A plane plate of homogeneous semi-conducting material lying horizontally is considered in this article (Fig. 1). The thickness and the width of the plate are denoted with $a$ and $b$, respectively. Along the plate length the axis OX of the coordinate system is drawn. The left and the right active ends of the plate have coordinates $x = 0$ and $x = L$, respectively.

The left and the right ends of the plate are connected via ohmic contact of a conductor with external electric circuit, which has a source of voltage, and causes the electrical potentials of plate ends.

The semi-conducting plate is illuminated from above with monochromatic light with wave length $\lambda$ and intensity $I_\lambda$. We suppose that the light intensity is constant all over the whole illuminated surface. At such geometry of the problem all parameters of problem depend only on the coordinate $x$.

In the considered one-dimensional case the electric field is constant in the material volume as is pointed in the negative direction of axis OX, and has a magnitude $E$. The drifted velocity is also equal through the whole volume.

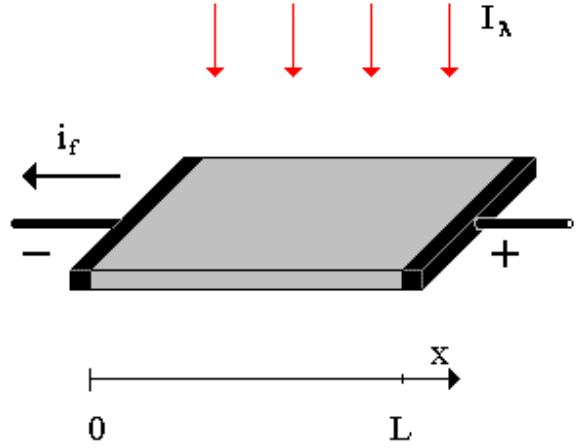

**Fig. 1. A special disposition of the semi-conducting material**

### 3.2 Equation of the Balance
At the plane (one-dimensional) and stationary case the equation (8) is reduced to

$$\frac{d}{dx}(N.\mu.E) = +k_\lambda.I_\lambda - \frac{N}{\tau}. \tag{9}$$

The equation (9) represents non-homogeneous linear differential equation of first order with constant coefficients for $N(x)$.

### 3.3 Solution of the Equation
Applying the standard procedure for the solving of equation (9), and after taking into account the limiting conditions we get

$$N(x) = k_\lambda.I_\lambda.\tau.\left[1 - exp\left(-\frac{x}{\tau.\mu.E}\right)\right]. \tag{10}$$

From where for the photo-current in material volume from (7) and (10) we will get

$$j_f(x) = q.k_\lambda.I_\lambda.\tau.\mu.E.\left[1 - exp\left(-\frac{x}{\tau.\mu.E}\right)\right]. \tag{11}$$

### 4. Analysis of the Solution
The analysis of the solution we will make as at first define some characteristic quantities:
- maximum value of current carriers concentration

$$N_{sat} = k_\lambda.I_\lambda.\tau ; \tag{12}$$

- photo-current of saturation $j_{sat}$

$$j_{sat} = q.k_\lambda.I_\lambda.L ; \tag{13}$$

- measure of intensity of the electric field $E_l$

$$E_I = \frac{L}{\tau \cdot \mu}.$$ (14)

Using (12) and (14) the expression (10) is transformed in the form

$$\frac{N(x)}{N_{sat}} = \left[1 - exp\left(-\frac{x}{L} \cdot \frac{E_I}{E}\right)\right].$$ (15)

The graph of (15) for three different values $E/E_I$ is presented in Fig. 2. As it is seen at weak fields the concentration of photo electrons is close to the magnitude of saturation (12) in a great part of material volume. With the increase of the field the concentration of photo electrons strongly decreases.

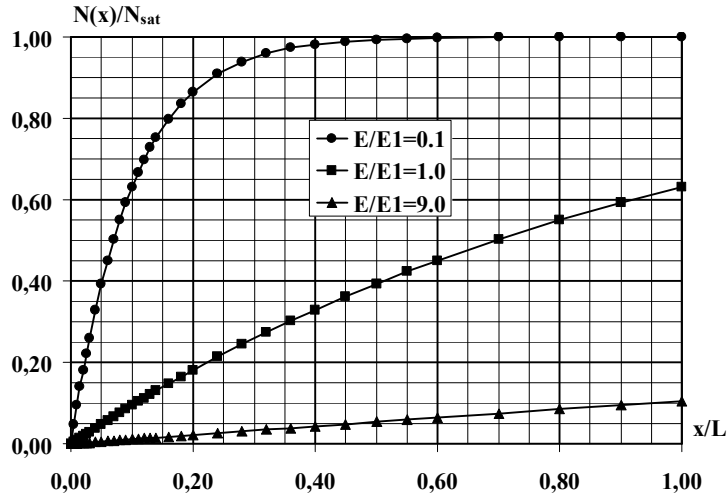

**Fig. 2. A graphic dependence of the concentration distribution of photo electrons in the material volume in accordance with the electric field intensity**

For the photo-current in material volume using (13) and (14) from (11) we will get

$$\frac{j_f(x)}{j_{sat}} = \frac{E}{E_I} \cdot \left[1 - exp\left(-\frac{x}{L} \cdot \frac{E_I}{E}\right)\right].$$ (16)

The graph of expression (16) for the same three different values $E/E_I$ is presented in Fig. 3. From the photo-current graph is seen that at weak fields the magnitude of photo-current in the volume does not undergo great changes, while at strong fields a considerable increase of the photo-current in the volume is available as the curve at very strong fields is close to straight line. Besides this makes sense to be marked the similarity of the dynamics between the concentration of photo electrons and photo-current, even though it occurs with different scales.

The physical interpretation which we can make is that we observe a competition between the processes of generation and loss of photo electrons with the participation of the external electric field. The external electric field orientates the photo electrons to the right, and brings them out of the material before they succeed to recombine in material volume. That's why with the raising of the field the photo-current increases.

At the positive end of the plate ($x = L$) the photo-current of photo electrons which leaves the material according (16) has a density as is following

$$\frac{j_f}{j_{sat}} = \frac{E}{E_I} \cdot \left[1 - exp\left(-\frac{E_I}{E}\right)\right].$$ (17)

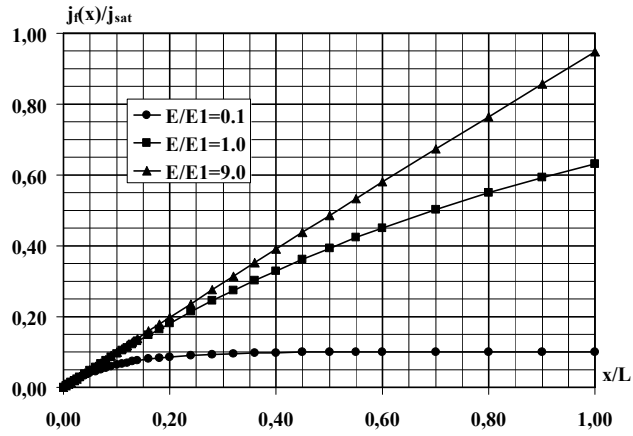

**Fig. 3. A graphic dependence of the distribution of phot-current of photo electrons in the material volume in accordance with the electric field intensity**

The graph of expression (17) is presented in Fig. 4.

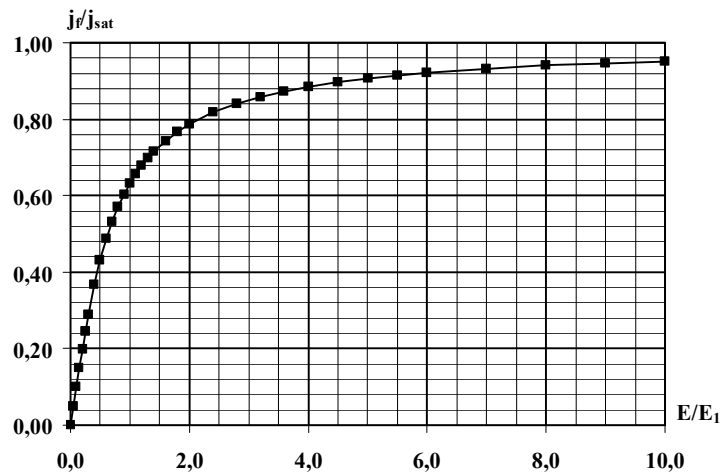

**Fig. 4. A graph of current-voltage characteristic of photo resistance**

As it is seen in Fig. 4 at weak fields a linear increase of photo-current is available, as with the raising of the electric field intensity it slows down its increase. But the photo-current can rise only up to the current of saturation (12), because at it all generated in the volume photo electrons are brought out of material volume. Thus, from the view of conception the presence of current of saturation is in principal necessary for the theory of photo-current of semi-conductor.

At weak fields ($E \ll E_1$) the expression (17) has the following approximate analytical form

$$\frac{j_f}{j_{sat}} \cong \frac{E}{E_1}. \qquad (18)$$

At strong fields ($E \gg E_1$) the expression (17) has the following approximate analytical form

$$\frac{j_f}{j_{sat}} \cong 1 - \frac{1}{2} \cdot \frac{E_1}{E} + \frac{1}{6} \cdot \left(\frac{E_1}{E}\right)^2. \qquad (19)$$

In fact in the expression (17) we have got the analytical form of current-voltage characteristic of photo resistance.

### 5. Conclusion

Finally we may conclude that as a result of the analysis of the equation of the balance of particles we got in analytical form the current-voltage characteristic of photo-current of semi-conducting plate which is uniformly illuminated by light.

---

# ВОЛТ-АМПЕРНА ХАРАКТЕРИСТИКА НА ФОТО СЪПРОТИВЛЕНИЕ. ПЛОСЪК СЛУЧАЙ

Димитър Г. СТОЯНОВ

**Резюме:** *В работата се изследва формирането на поток от фотоелектрони в обема на полупроводников материал. Разглежда се плосък случай с равномерно осветяване на материала със светлина. Получена е волт-амперната характеристика на фото-съпротивление в аналитичен вид.*

**Ключови думи:** полупроводник, фотопроводимост, волт-амперна характеристика

# ВОЛТ-АМПЕРНА ХАРАКТЕРИСТИКА НА ФОТО СЪПРОТИВЛЕНИЕ. ПЛОСЪК СЛУЧАЙ

ДИМИТЪР СТОЯНОВ

**Резюме:** *В работата се изследва формирането на поток от фотоелектрони в обема на полупроводников материал. Разглежда се плосък случай с равномерно осветяване на материала със светлина. Получена е волт-амперната характеристика на фото-съпротивление в аналитичен вид.*

**Ключови думи:** полупроводник, фотопроводимост, волт-амперна характеристика

# CURRENT -VOLTAGE CHARACTERISTIC OF A PHOTO RESISTANSE. A PLANE CASE

DIMITAR STOYANOV

**Abstract:** *The formation of photo-electron current in the volume of semi-conductor material is investigated in this article. A plane case of material uniform lighting with light is considered. The current-voltage characteristic of a photo resistance is obtained in analytical form.*

**Key words:** *semi-conductor, photoconductivity, current-voltage characteristic*

## 1. УВОД

Фотопроводимостта е важна характеристика на полупроводниковите материали. В настоящият етап на развитие на електрониката фотопроцесите в полупроводниците и полупроводниковите детайли намират все по-голямо практическо приложение. Затова става все по-належаща необходимостта в учебниците [1, 3] да се увеличат обемите, предназначени за изучаването на тези процеси.

Целта в тази работа е да се направи анализ на процесите който участват във формирането на фототока и неговото протичане. Разглежда се плосък случай с равномерно осветяване на полупроводниковия материал.

## 2. БАЛАНС НА ЧАСТИЦИТЕ
### 2.1. Основни процеси

При осветяване на повърхността на полупроводников материал със светлина, тя прониква в обема на материала и в резултат на вътрешен фотоефект предизвиква появата на електрони в зоната на проводимостта на полупроводника. Тези електрони ще наричаме фотоелектрони, за да ги различаваме от електроните в зоната на проводимостта, които могат да бъдат от основните токоносители или от легиране на материала.

Генерацията на фотоелектрони $N$ в единица обем, за единица време е пропорционална на интензитета на монохроматичната светлина $I_\lambda$ [1, 2]. Коефициентът на пропорционалност $k_\lambda$ отчита спектралната чувствителност на материала по дължини на вълните $\lambda$ на падащата светлина

$$\left(\frac{\delta N}{\delta t}\right)_{ph} = +k_\lambda . I_\lambda \qquad (1)$$

Основният механизъм за загуба на фотоелектроните е рекомбинацията им с дупки $N_p$ от зоната на валентността. При малък интензитет на светлината, дупките генерирани от фотоефекта са много по-малко от дупките на основните и неосновните токоносители. Поради това голяма част от процеса на рекомбинация ще настъпи с дупки, концентрацията на които не зависи от светлината. Затова можем да въведем характерно време на живот на фотоелектроните $\tau$, което е постоянна, независеща от интензитета на светлината величина [1, 2].

$$\left(\frac{\delta N}{\delta t}\right)_{rec} = -\gamma.N.N_p = -\frac{N}{\tau}. \quad (2)$$

Поради малкият интензитет на светлината фотоелектроните няма да влияят на интензитета на електричното поле [1, 2] и ще се изпълнява законът на Гаус

$$div\vec{E} = 0. \quad (3)$$

### 2.2. Баланс на частиците

Балансът на фотоелектроните, по наше мнение, трябва да се направи, използвайки уравнението на непрекъснатостта, в дясната страна на което поставяме сумата на скоростите от (1) и (2)

$$\frac{\partial N}{\partial t} + div\vec{\Gamma} = +k_\lambda.I_\lambda - \frac{N}{\tau}. \quad (4)$$

Тук в (4) фигурира величината $\vec{\Gamma}$ - поток на фотоелектроните, породен от наличието на електрично поле с интензитет $\vec{E}$. Ще предполагаме, че скоростта на потока съвпада с дрейфовата скорост на фотоелектроните $\vec{u}_d$

$$\vec{u}_d = -\mu.\vec{E}. \quad (5)$$

В (5) с $\mu$ е означена подвижността на фотоелектроните в материала ($\mu > 0$). Самият поток на фотоелектроните ще има вида

$$\vec{\Gamma} = N.\vec{u}_d = -N.\mu.\vec{E}. \quad (6)$$

Потокът фотоелектрони ще формира ток на фотоелектроните (който ще наричаме фототок) с плътност $\vec{j}_f$

$$\vec{j}_f = -q.\vec{\Gamma} = q.N.\mu.\vec{E}. \quad (7)$$

Тук $q$ е големината на заряда на електрона ($q > 0$). Замествайки (6) в (4) ще получим

$$\frac{\partial N}{\partial t} + div\left(-N.\mu.\vec{E}\right) = +k_\lambda.I_\lambda - \frac{N}{\tau}. \quad (8)$$

Това е уравнението на баланса на фотоелектроните в най-общ случай. То се решава съвместно с (3) при следните гранични условия:
а) Върху отрицателният край на материала концентрацията и потока на фотоелектроните е нула;
б) Върху положителния край на материала концентрацията и потока на фотоелектроните е максимална. Потока на фотоелектроните задава фототока $i_f$, регистриран през материала.

### 3. ПЛОСЪК ОБРАЗЕЦ
#### 3.1. Геометрия на задачата

В настоящата работа разглеждаме плоска пластинка от хомогенен полупроводников материал, лежаща хоризонтално (виж Фиг.1). Дебелината на пластинката е a, а ширината b. По дължината на пластинката е прекарана оста OX на координатната система. Левият активен край на пластинката има координата $x = 0$, а десният активен край на пластинката има координата $x = L$.

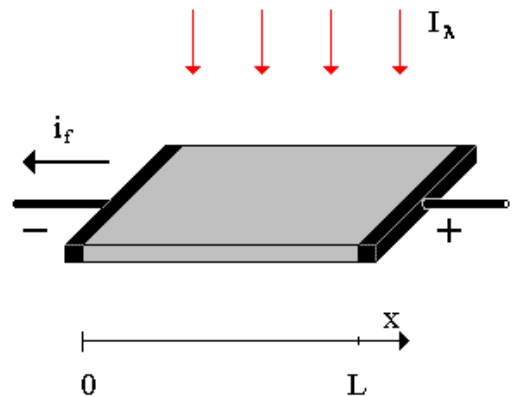

***Фиг1.*** *Пространствено разположение на полупроводниковия материал*

Левият и десният край на пластинката, чрез оммичен контакт с проводник се свързва с външна електрическа верига, която притежава източник на напрежение и задава електрическият поляритет на краищата на пластинката.

Полупроводниковата пластинка се осветява отгоре с монохроматична светлина с дължина на вълната $\lambda$ и интензитет $I_\lambda$. При това разглеждане предполагаме, че интензитета на светлината е постоянен по цялата осветявана повърхност. При такава геометрия на задачата всички параметри на задачата могат да зависят единствено от координатата $x$.

В разглеждания едномерен случай електричното поле е постоянно в обема на материала, насочено е в отрицателната посока на оста ОХ и има големина $E$. Дрейфовата скорост също е еднаква в целия обем.

### 3.2. Уравнение на баланса
В плоския (едномерен) и стационарен случай уравнението (8) се редуцира до

$$\frac{d}{dx}(N.\mu.E) = +k_\lambda.I_\lambda - \frac{N}{\tau}. \qquad (9)$$

Уравнение (9) представлява нехомогенно линейно диференциално уравнение от първи род с постоянни коефициенти за $N(x)$.

### 3.3. Решение на уравнението
Прилагайки стандартната процедура за решаване на (9), след отчитане на граничните условия получаваме

$$N(x) = k_\lambda.I_\lambda.\tau\left[1 - exp\left(-\frac{x}{\tau.\mu.E}\right)\right]. \qquad (10)$$

Откъдето за фототока в обема на материала от (7) и (10) ще получим

$$j_f(x) = q.k_\lambda.I_\lambda.\tau.\mu.E\left[1 - exp\left(-\frac{x}{\tau.\mu.E}\right)\right]. \qquad (11)$$

### 4. АНАЛИЗ НА РЕШЕНИЕТО
Анализът на решението ще направим, като първо дефинираме няколко характерни величини:

- максималната стойност на концентрацията на токоносителите

$$N_{sat} = k_\lambda.I_\lambda.\tau; \qquad (12)$$

- фототок на насищане $j_{sat}$

$$j_{sat} = q.k_\lambda.I_\lambda.L; \qquad (13)$$

- мащаб за интензитета на електричното поле $E_I$

$$E_I = \frac{L}{\tau.\mu}. \qquad (14)$$

Ползвайки (12) и (14), изразът (10) се преобразува във вида

$$\frac{N(x)}{N_{sat}} = \left[1 - exp\left(-\frac{x}{L}.\frac{E_I}{E}\right)\right]. \qquad (15)$$

На Фиг.2 е дадена графиката на (15) за три различни стойности $E/E_I$. От фигурата се вижда, че при слаби полета, концентрацията на фотоелектроните е близка до стойността на насищане (12) в голяма част от обема на материала. С увеличаване на полето концентрацията на фотоелектроните силно намалява.

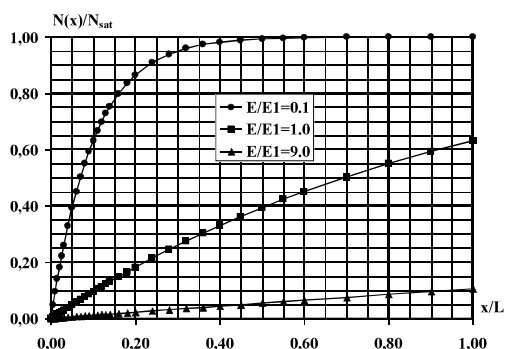

**Фиг. 2.** Графика на разпределението на концентрацията на фотоелектроните в обема на материала в зависимост от интензитета *на електричното поле*

За фототока в обема на материала, ползвайки (13) и (14), от (11) ще получим

$$\frac{j_f(x)}{j_{sat}} = \frac{E}{E_I}.\left[1 - exp\left(-\frac{x}{L}.\frac{E_I}{E}\right)\right]. \qquad (16)$$

На Фиг.3 е дадена графиката на (16) за същите три различни стойности $E/E_I$. От графиката за фототока Фиг.3 се вижда, че при слаби полета големината на фототока в обема не търпи големи промени. Докато при силни полета имаме силно нарастване на фототока в обема, като кривата при много силните полета е близка до права линия. Освен това има смисъл да се отбележи подобието на динамиката между концентрацията на фотоелектроните и фототока, макар и това да става с различни мащаби.

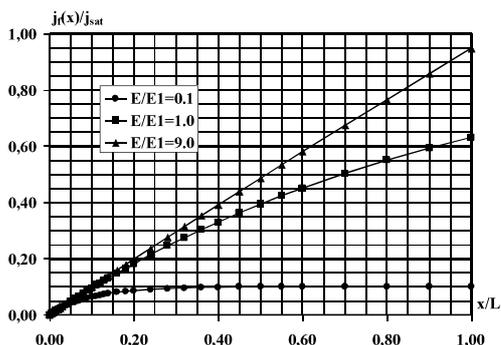

**Фиг. 3.** *Графика на разпределението на фототока на фотоелектроните в обема на материала в зависимост от интензитета на електричното поле*

Физическото тълкуване, което можем да направим е, че наблюдаваме конкуренция между процесите на генерация и загуба на фотоелектроните, с участие на външното електрично поле. Външното електрично поле насочва фотоелектроните надясно и ги извежда от материала, преди да успеят да рекомбинират в обема на материала. Затова с увеличаване на полето фототока расте.

На положителния край на пластината ($x = L$) фототока на фотоелектроните, който напуска материала, според (16) има плътност

$$\frac{j_f}{j_{sat}} = \frac{E}{E_1} \cdot \left[ 1 - \exp\left(-\frac{E_1}{E}\right) \right]. \qquad (17)$$

Графиката на зависимостта (17) е дадена на Фиг.4.

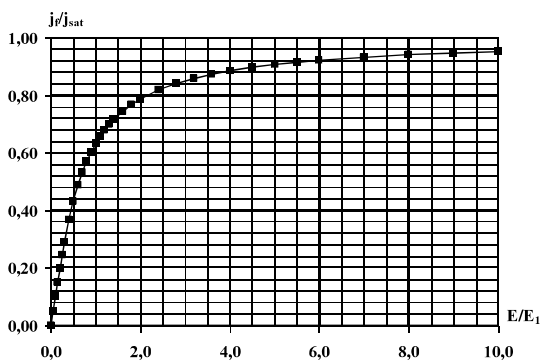

**Фиг.4** *Графика на волт-амперната характеристика на фотосъпротивление*

Както се вижда от фигурата, при слаби полета имаме линейно нарастване на фототока, който с учеличаване на интензитета на електричното поле забавя своето нарастване. Но фототока може да нарасне само до тока на насищане (12), защото при него всички генерирани в обема фотоелектрони са изведени извън обема на материала. Така, че концептуално погледнато наличието на ток на насищане е принципно необходимо за теорията на фототока на полупроводник.

При слаби полета ($E \ll E_1$) зависимостта (17) има следния приблизителен аналитичен вид:

$$\frac{j_f}{j_{sat}} \cong \frac{E}{E_1}. \qquad (18)$$

При силни полета ($E \gg E_1$) зависимостта (17) има следния приблизителен аналитичен вид:

$$\frac{j_f}{j_{sat}} \cong 1 - \frac{1}{2} \cdot \frac{E_1}{E} + \frac{1}{6} \left(\frac{E_1}{E}\right)^2. \qquad (19)$$

Всъщност, в лицето на (17) получихме аналитичния вид на волт-амперната характеристика на фотосъпротивление.

## 5. ЗАКЛЮЧЕНИЕ

Накрая можем да кажем, че в резултат на анализ на уравнението на баланс на частиците е получена в аналитичен вид волт-амперната характеристика на фототока на полупроводникова пластина с равномерно осветяване.

## ЛИТЕРАТУРА

Department of Mathematics, Physics and Chemistry
Technical University–Sofia, EPF - Sliven
59 Bourgasko shosse Blv
8800 Sliven
BULGARIA
E-mail: dgstoyanov_ipf@abv.bg